\documentclass[twocolumn,10pt]{article}

\title{Classifying vortex wakes using neural networks}
\author{Brendan Colvert, Mohamad Alsalman, and Eva Kanso\footnote{Corresponding author: Kanso@usc.edu} \\
\small{Aerospace and Mechanical Engineering, University of Southern California, Los Angeles, CA 90089}}
\date{}

\usepackage[margin=1in]{geometry}
\usepackage{amsmath}
\usepackage{gensymb}
\usepackage{graphicx}
\usepackage{bm}
\usepackage{tikz}
\usepackage{amsfonts}
\usepackage{booktabs}
\usepackage{multirow}
\usepackage{fourier}
\usepackage{upgreek}
\usepackage{xcolor}

\begin{document}
\maketitle

\begin{abstract}
Unsteady flows  contain information about the objects creating them.
Aquatic organisms offer intriguing paradigms for 
extracting flow information using local sensory measurements.
In contrast, classical methods for flow analysis require global knowledge of the flow field. 
Here, we train neural networks to classify flow patterns using local vorticity measurements.
Specifically, we consider vortex wakes behind an oscillating airfoil and we evaluate the accuracy of the network in distinguishing between three wake types, 2S, 2P+2S and 2P+4S. The network uncovers the salient features of each wake type. 
\end{abstract}

\section{Introduction}

Objects moving in a fluid medium often leave behind long-lived vortical flows. These flows contain distinct hydrodynamic 
cues that can, in principle, be detected and even exploited for navigation and motion planning~\cite{Spedding2013}.  
Detection of hydrodynamic signals is well documented in aquatic organisms at various length and time scales.
The organism ``feels" the surrounding fluid through specialized sensory modalities such as pressure or velocity
sensors and responds accordingly; see, e.g.,~\cite{Colvert2017a,Colvert2017b} and references therein. For example, harbor seals are known to track a moving object by following its wake~\cite{Dehnhardt2001}, and fish respond to a wide variety of flow stimuli in behaviors ranging from 
upstream migration to predator evasion~\cite{Colvert2016,Engelmann2000,Montgomery1997,Ristroph2015}.  
One of the most fascinating aspects of these behaviors, and the main inspiration for the present study, is the ability of the organism to distinguish between different flow patterns by relying on local flow information only.
Here, we use neural networks to investigate how certain wake patterns can be identified and classified locally, with no knowledge of the global flow field.

\begin{figure}[t!]
\centering
\includegraphics{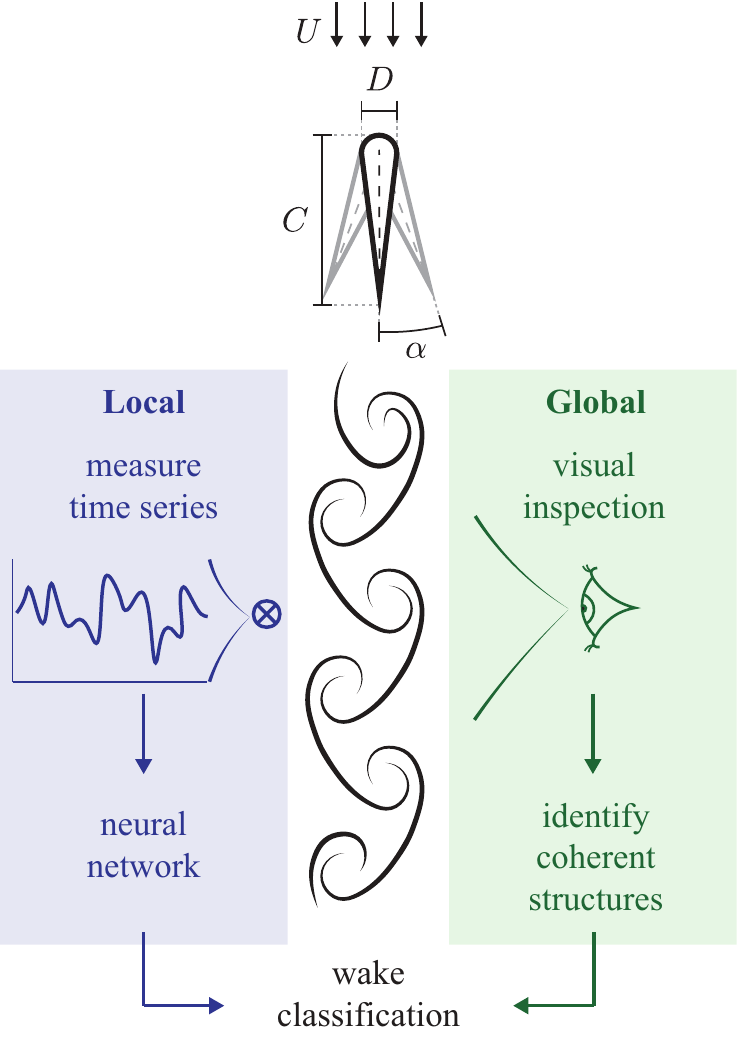}
\caption{%
Classification of flow patterns behind an oscillating airfoil: (left) proposed approach using only local measurements, and (right) classic approach relying on global inspection of the spatiotemporal patterns in the flow field.
}
\label{fig:global_local}
\end{figure}

\begin{figure}[t!]
\centering
\includegraphics{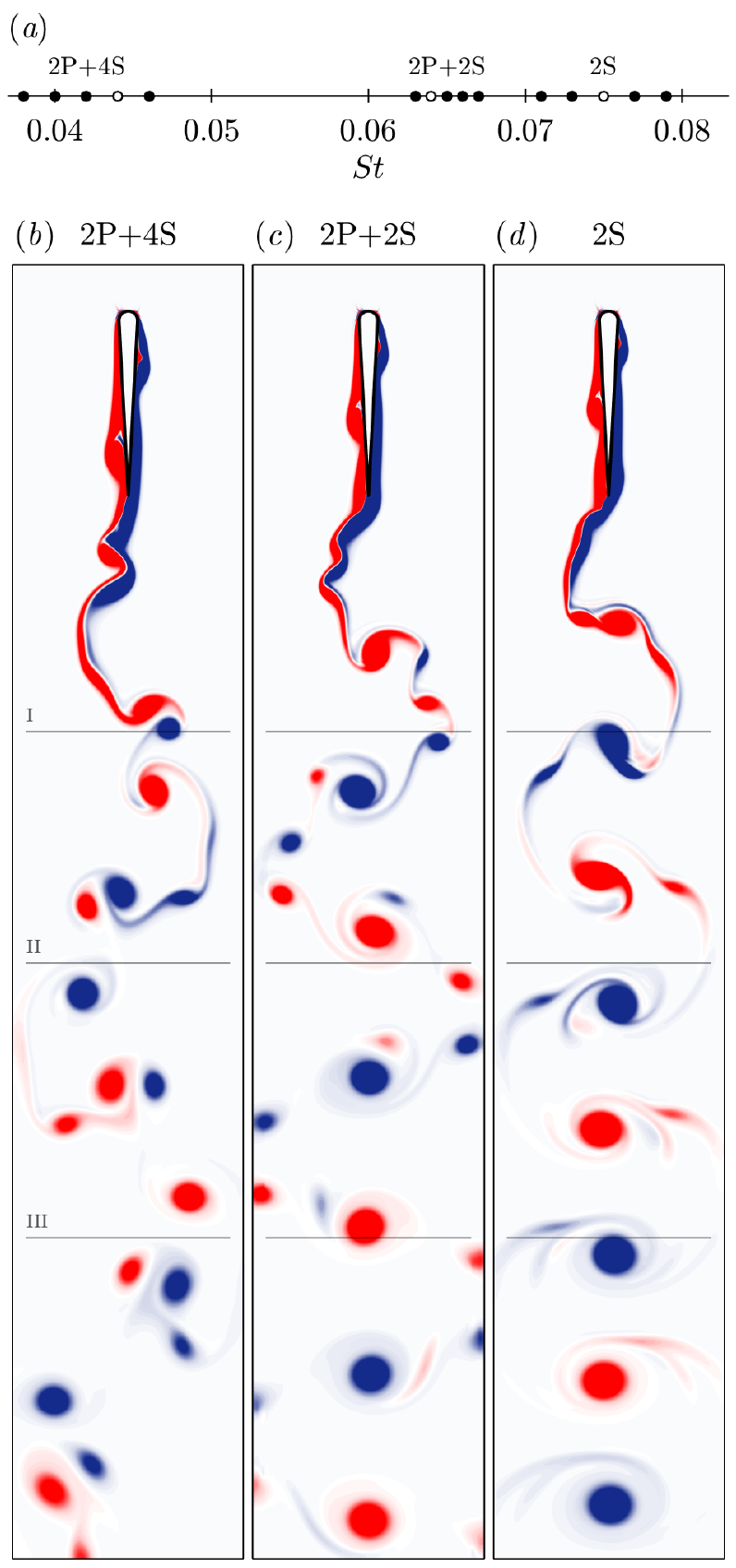}
\caption{%
Wake behind an oscillating airfoil for $\text{A} = 1.656$ and  St ranging from $0.038$ to $0.079$ as shown in (a). 
Three wake types are observed, 2P+4S,  2P+2S, and 2S, shown respectively by plotting contours of vorticity field $\omega({\bm{x}},t)$ for (b) $\text{St}~=~0.044$, (c) $\text{St}~=~0.064$ and (d) $\text{St}~=~0.075$. }
\label{fig:wakes}
\end{figure}

The spatiotemporal organization of vorticity in the unsteady wake of a moving body is intimately linked to the parameters of the body and its motion. Classical examples include the von K\'{a}rm\'{a}n vortex street behind a stationary cylinder in a uniform free stream  and the more ``exotic" vortex wakes  behind oscillating cylinders~\cite{Williamson1988} and  airfoils~\cite{Bratt1953, Buchholz2008, Godoy2008, Koochesfahani1986, Koochesfahani1989, Lai1999,  Schnipper2009},
as well as the wake structures of swimming fish~\cite{Muller2008, Tytell2004,Ysasi2011} and flapping wings~\cite{Platzer2008}. 
Vortical wakes are often described by the number of vortices shed per oscillation or flapping period~\cite{Williamson1988}; 2S  refers to wakes in which two vortices of opposite sign are shed per period whereas in 2P wakes, two vortex pairs  are shed per period.
Classifying the wake type (2S, 2P, etc.) as a function of the body parameters is relevant to many applications including the design of offshore structures and the analysis of propulsive forces in aquatic and aerial organisms and bio-inspired vehicles. 
The classical approach to wake classification relies on global inspection of the flow field by an external observer in order to identify global topological patterns \cite{Schnipper2009, Stremler2014,Williamson1988} or Lagrangian coherent structures \cite{Haller2005,Jeong1995}. This global approach provides valuable insight for engineering design and analysis but it is not suited for applications where one has access to local information only. The objective of this study
is to classify global flow patterns using local flow measurements.

To address this problem, we use supervised machine learning and neural networks.
Neural networks provide versatile and powerful tools that have been applied to many branches of engineering and science ranging from autonomous vehicles \cite{Dahlkamp2006} to skin cancer classification \cite{Esteva2017}.
This versatility is due to the fact that neural networks, with as few as one hidden layer, can approximate any continuous function with arbitrary accuracy \cite{Hornik1989}. In addition, recent advances in computational capabilities and fast algorithms  \cite{Hecht-Nielsen1989,Werbos1990} made it relatively straighforward to design, construct and train neural networks. Yet, the use of neural networks, and more generally machine learning algorithms, to solve problems in fluid mechanics is sparse. Exceptions include the use
of neural networks for flow modeling~\cite{Lee1997,Muller1999} and reinforcement learning algorithms in bioinspired locomotion and decision making~\cite{Gazzola2014, Gazzola2016, Novati2017}. In this paper, we design a neural network that classifies the wake type from local time measurements of the vorticity field in the wake of an oscillating airfoil; see figure~\ref{fig:global_local}. Specifically, we train the neural network to distinguish between three wake types: 2S, 2P+2S, and 2P+4S. We analyze the performance of the trained network in terms of its accuracy in predicting the wakes of test simulations not used for training and we discuss the salient features relevant to classify each wake type.

\begin{figure*}
\centering
\includegraphics{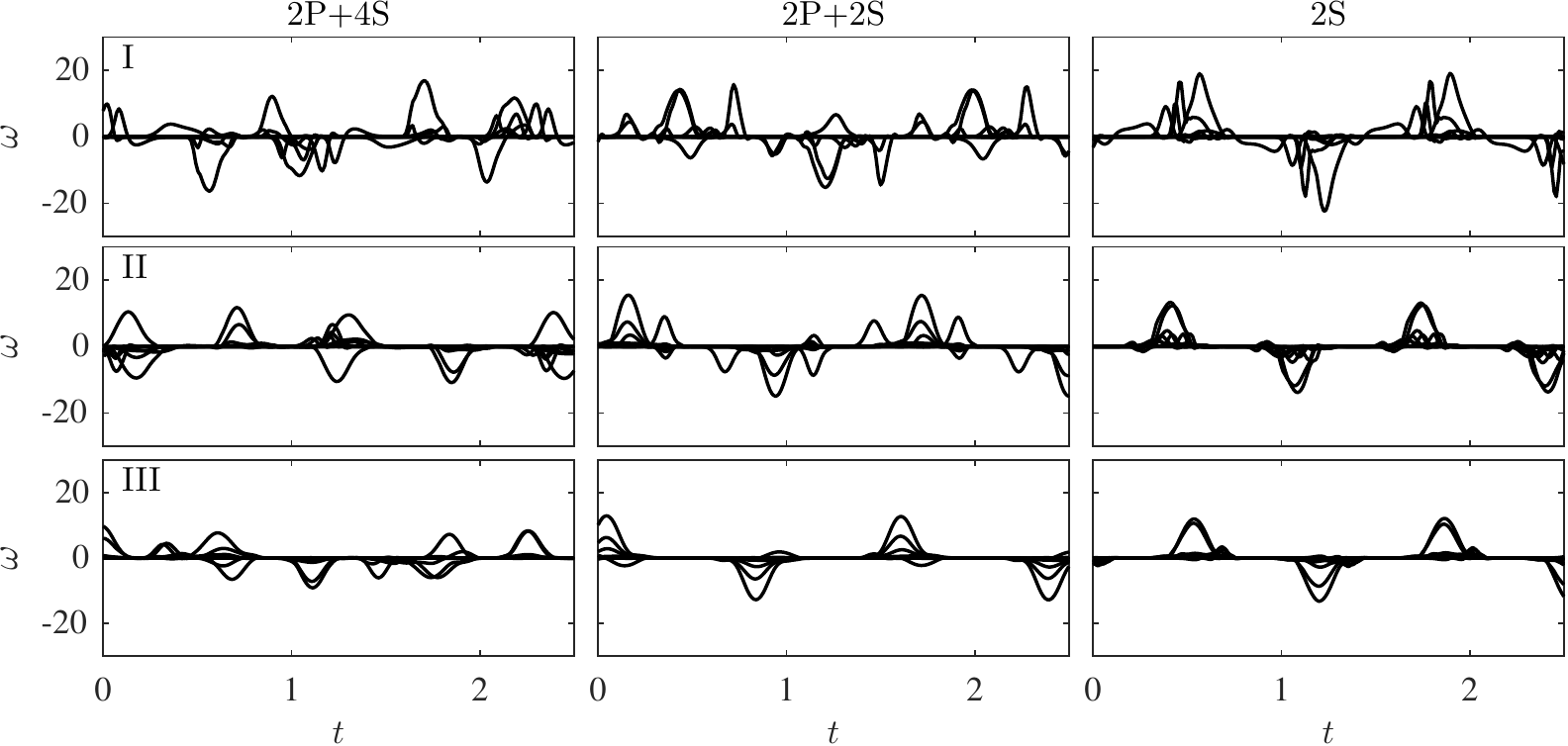}
\caption{%
Time traces of vorticity in the wake of an oscillating airfoil for wake types 2P+4S, 2P+2S, and 2S at select locations I, II and III highlighted in figure~\ref{fig:wakes}.
}
\label{fig:time_traces}
\end{figure*}

\section{Methods}

\subsection{Vortex wakes of oscillating airfoils}

We consider a symmetric airfoil of chord $C$ and diameter $D$ undergoing a simple pitching motion of the form
\begin{equation}
\phi(t) = \alpha \sin 2\pi f t,
\end{equation}
where $\phi$ is the angle of attack, $\alpha$ is the maximum angle of the airfoil, $f$ is the frequency of oscillation, and $t$ is time;  see figure~\ref{fig:global_local}.
This system can be characterized by three dimensionless parameters: the Strouhal number St, which is a dimensionless measure of the oscillation frequency, the dimensionless amplitude A of the airfoil trailing edge, and the Reynold number Re,
\begin{equation}
\text{St} = \frac{f D}{U}, \quad \text{A} = \frac{2C\sin \alpha}{D},  \quad \text{Re} = \dfrac{\rho U D}{\mu} .
\end{equation}
Here, $\rho$ and $\mu$ are the fluid density and dynamic viscosity, and $U$ is the velocity of the uniform free stream.
For Reynolds numbers $\text{Re}$ between $10^2$ and $10^4$, the qualitative structure of the wake is independent of Re and depends on the Strouhal number St and oscillation amplitude A~\cite{Schnipper2009}. 
The Strouhal number is inversely proportional to the oscillation period, which is typically an integer multiple of the vortex shedding period. 
At higher $\text{St}$, the number of vortices shed per period is generally two, a set of oppositely-signed vortices, and the wake type is 2S.
As $\text{St}$ decreases, the number of vortices shed per period increases and more exotic wakes are observed, including 2P+2S and 2P+4S wakes~\cite{Schnipper2009}.

To obtain numerical data of the wake, we  solve the incompressible Navier-Stokes equations,
\begin{equation}
\dfrac{\partial\bm{u}}{\partial t} + \bm{u}\cdot \nabla \bm{u} = - \nabla p + \dfrac{1}{\text{Re}} \Delta \bm{u}, \quad \nabla\cdot \bm{u} = 0,
\end{equation}
 in the fluid domain surrounding the oscillating airfoil. Here, 
 the fluid velocity field $\bm{u}$ and the pressure field $p$ depend on space, parameterized by the position vector $\bm{x}$, and time $t$. Further, $\bm{u}$ is subject to the no-slip condition at the airfoil boundary. Since the airfoil motion is prescribed, the  fluid-structure interactions are one-way coupled: the airfoil affects the flow but not vice-versa. To solve for $\bm{u}(\bm{x},t)$, we use a numerical algorithm based on the Immersed Boundary Method, initially devised for fully-coupled  fluid-structure interactions~\cite{Peskin1977}. The algorithm is described in detail in~\cite{Mittal2008}, and has been implemented, optimized and tested extensively by the group of Haibo Dong; see, e.g., \cite{Vargas2008, Bozkurttas2009}.  We then calculate the vorticity field $\bm{\omega} = \nabla \times \bm{u}$, which can be expressed as a scalar field $\omega(\bm{x},t)$ given the two-dimensional nature of the problem. 
 
 We conducted a total of fifteen flow simulations by varying $\text{St}$ from $0.038$ to $0.079$ 
 while fixing $\text{A} = 1.656$ and $\text{Re} = 500$. 
Three different types of periodic wakes are observed: 2P+4S, 2P+2S, and 2S, each occurring at five distinct values of $\text{St}$ 
as shown in figure~\ref{fig:wakes}(a). These wake types are identified by global inspection of spatiotemporal patterns of the 
vorticity field $\omega$. Representative examples of these wakes are shown by plotting the contours of the vorticity field $\omega$ 
in figures~\ref{fig:wakes}(b)-(d)

\begin{figure*}[!t]
\centering
\includegraphics{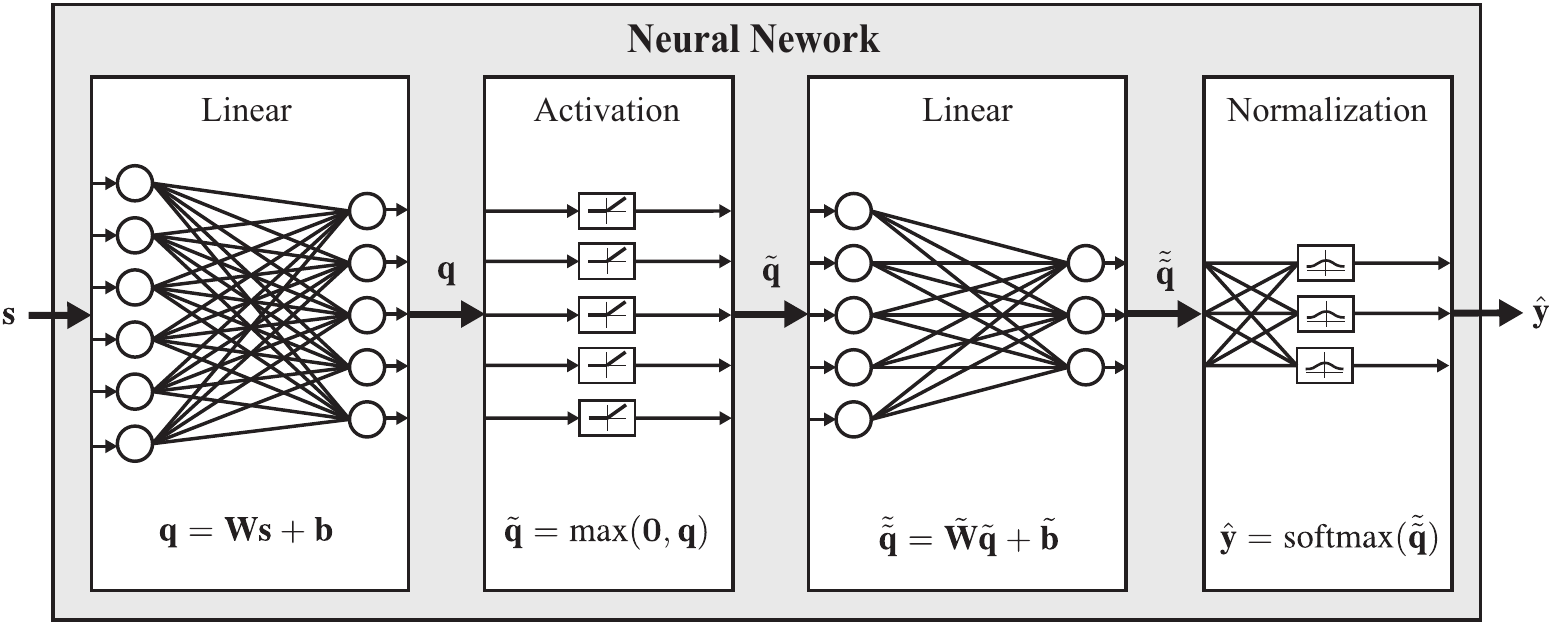}
\caption{%
Architecture of the neural network
consisting of four successive layers: a linear layer,  a nonlinear activation layer, another linear layer, and a nonlinear normalization layer.
}
\label{fig:neural_network}
\end{figure*}

\subsection{Local flow measurements}

We measure the vorticity field $\omega(\bm{x},t)$ at a position $\bm{x}_{m}$ in the wake using a sampling
interval $\Delta t$.  Examples of the time evolution of the local vorticity are shown in figure~\ref{fig:time_traces}. 
We represent the local vorticity measurements as a column vector  $\textbf{s}_m$ of length $N$, where $N$ is the total number of samples. In other words, the {$n^{\text{th}}$} entry of the measurement vector $\textbf{s}_m$ is equal to $\omega(\bm{x}_m,n\Delta t)$. 
Our goal is to train a neural network that takes as input the measurement vector $\textbf{s}_m$, also called a {\em snapshot}, and classifies the wake type into 2P+4S, 2P+2S or 2S.

We construct a training dataset $\textbf{S}$ $\in \mathbb{R}^{N \times M}$ of $M$ snapshots
\begin{equation}
\textbf{S} = \begin{bmatrix}
| & | & & |\\
\textbf{s}_1 & \textbf{s}_2 & \cdots & \textbf{s}_M\\
| & | & & |
\end{bmatrix}.
\label{eq:data_matrix}
\end{equation}
We then assign to each snapshot $\textbf{s}_m$ a  \emph{label} $\textbf{y}_m$ that identifies the wake type or \emph{class}.
The three classes 2P+ 4S, 2P+ 2S, {and} 2S are represented, respectively, by
\begin{equation}
\textbf{y}_m = \begin{bmatrix}
1\\
0\\
0
\end{bmatrix}, \quad
\begin{bmatrix}
0\\
1\\
0
\end{bmatrix}, \quad  \textrm{and} \quad
 \begin{bmatrix}
0\\
0\\
1
\end{bmatrix}.
\label{eq:outputclasses}
\end{equation}
Here, $\textbf{y}_m$ is  a column vector of length $N_c = 3$, where $N_c$ is the number of classes.
The class label matrix $\textbf{Y} \in \mathbb{R}^{N_c \times M}$ for the training set can be written as
\begin{equation}
\textbf{Y} = \begin{bmatrix}
| & | & & |\\
\textbf{y}_1 & \textbf{y}_2 & \cdots & \textbf{y}_M\\
| & | & & |
\end{bmatrix}.
\end{equation}

\begin{figure*}
\centering
\includegraphics{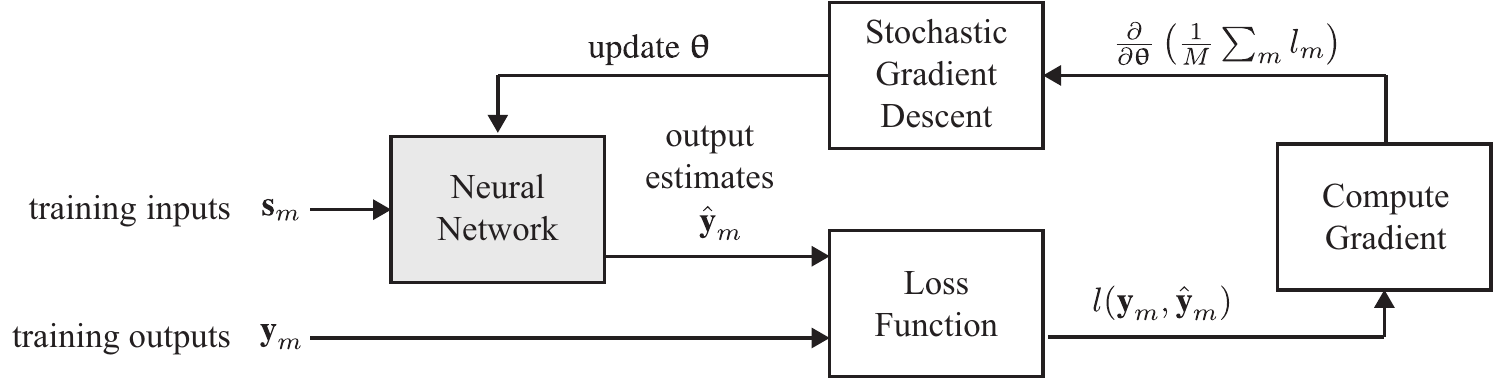}
\caption{%
A neural network is trained by optimizing a loss function over the network parameters $\bm{\uptheta}$ subject to a training set of snapshots $\textbf{s}_m$ and labels $\textbf{y}_m$. The optimization problem is solved using stochastic gradient descent.
}
\label{fig:training}
\end{figure*}

For each flow simulation, we probe the wake at select locations $\bm{x}_m$ chosen according to an equally-spaced grid of 51$\times$121 points. 
Ten snapshots are recorded at each grid point as follows.  First, the time series of the local vorticity field is
sampled at $\Delta t = 0.012$ for $N=210$, resulting in a total time of approximately $2.5$ units.
We then shift the time at which the first sample is recorded by about $0.25$ time units to create a total of ten distinct snapshots.
That is to say, from each flow simulation, we extract a total of 51$\times$121$\times$10 = 61,710 snapshots, resulting in 15$\times$61,710 = 925,650 snapshots from all simulations.

\subsection{Neural networks}
\label{sec:network}

A {neural network} is formally defined  as a mapping 
$\hat{\textbf{y}} = f(\textbf{s},\bm{\uptheta})$ 
from an input $\textbf{s} \in  \mathbb{R}^{N}$ to an estimate 
$\hat{\textbf{y}} \in \mathbb{R}^{N_c}$ of the class type. Here, $\bm{\uptheta}$ denotes the mapping parameters. Note that the hat is used to distinguish the estimate $\hat{\textbf{y}}$ from the actual label ${\textbf{y}}$. Also note that we temporarily drop the subscript {$(\cdot)_m$} for clarity.
The mapping $f$ is constructed by the successive application of various functions, each one is called a \emph{layer}.
The particular functional form for each layer and the number of layers depend on the specific application~\cite{Schmidhuber2015}. 
In this work, we use the feedforward neural network shown in figure~\ref{fig:neural_network}.

\paragraph{Network architecture} We design a feedforward neural network of four layers. The first layer is an affine transformation
\begin{equation}
\textbf{q} = \textbf{W}\, \textbf{s} + \textbf{b},
\label{eq:layer_1}
\end{equation}
where $\textbf{q}\in \mathbb{R}^{N_f}$ is the output vector, $\textbf{W}  \in \mathbb{R}^{N_f \times N}$ is the \emph{weight matrix}, $\textbf{b} \in \mathbb{R}^{N_f}$ is the \emph{bias vector}, and $N_f$ is the number of features.
We interpret the rows of $\textbf{W}$ as features of the vorticity time series that are relevant for differentiating between classes. Consequently, $\textbf{W}\textbf{s}$ is a projection of the 
input $\textbf{s}$ onto these features. 

The second layer is called an \emph{activation layer} and is defined by the nonlinear function $\tilde{\textbf{q}} = \text{max} \hspace{1pt} (\bm{0},\textbf{q})$
that sets negative entries of $\textbf{q}$ to zero and `activates' only positive entries. In component form, one has
\begin{equation}
\tilde{\text{q}}_j = \begin{cases}
\text{q}_j & \text{q}_j \geq 0,\\
0 & \text{q}_j < 0,
\end{cases} \qquad j = 1, \ldots, N_f.
\label{eq:layer_2}
\end{equation}
The third layer is another affine transformation 
\begin{equation}
\tilde{\tilde{\textbf{q}}} = \tilde{\textbf{W}}\, \, \tilde{\textbf{q}} + \tilde{\textbf{b}},
\label{eq:layer_3}
\end{equation}
where $\tilde{\tilde{\textbf{q}}} \in \mathbb{R}^{N_c}$, $\tilde{\textbf{W}} \in \mathbb{R}^{N_c \times N_f}$, and $\tilde{\textbf{b}} \in \mathbb{R}^{N_c}$.
Finally, the last layer is a \emph{normalization layer} $\hat{\textbf{y}} = \text{softmax} (\tilde{\tilde{\textbf{q}}})$, which can be written in component form as  
\begin{equation}
 \hat{\text{y}}_c = \dfrac{ \exp (\tilde{\tilde{\text{q}}}_c)}{ \sum_{c} \exp(\tilde{\tilde{\text{q}}}_c)}, \qquad  c = 1, \ldots, N_c.
\label{eq:layer_4}
\end{equation}
This layer normalizes the output vector $\hat{\textbf{y}}$ such that its entries  take values between 0 and 1 and its $L^1$-norm $\| \hat{\textbf{y}} \| = 1$. 
The output $\hat{\textbf{y}}$  is a probability distribution vector, where the $c^{\textrm{th}}$ entry $\hat{\text{y}}_c$ represents the likelihood that $\textbf{s}$ belongs to class $c$.

The mapping $\hat{\textbf{y}} = f(\textbf{s},\bm{\uptheta})$ is constructed by successive application of 
these four layers. The parameter vector $\bm{\uptheta}$ consists of all the entries of the weight matrices $\textbf{W}$ and $\tilde{\textbf{W}}$ and bias vectors $\textbf{b}$ and $\tilde{\textbf{b}}$. That is, $\bm{\uptheta} \in \mathbb{R}^{N_p}$ is a column vector of dimension
\begin{equation}
N_p = N_f(N_c+N+1) + N_c.
\label{eq:num_params}
\end{equation}

\paragraph{Network training}

The \emph{training} algorithm, depicted schematically in figure~\ref{fig:training}, optimizes the values of 
 $\bm{\uptheta}$ subject to a loss function of the form
\begin{equation}
l(\textbf{y} , \hat{\textbf{y}}) = \sum_{c=1}^{N_c} -\text{y}_{c} \ln \hat{\text{y}}_{c} - (1-\text{y}_{c}) \ln ( 1- \hat{\text{y}}_{c} ).
\label{eq:loss}
\end{equation}
This loss function is zero only when $\hat{\textbf{y}} = \textbf{y}$ and nonnegative otherwise. 
The logarithmic form of the function means that the loss  increases dramatically as the estimate $\hat{\textbf{y}}$ deviates further from
$\textbf{y}$.

\begin{figure*}[!t]
\centering
\includegraphics{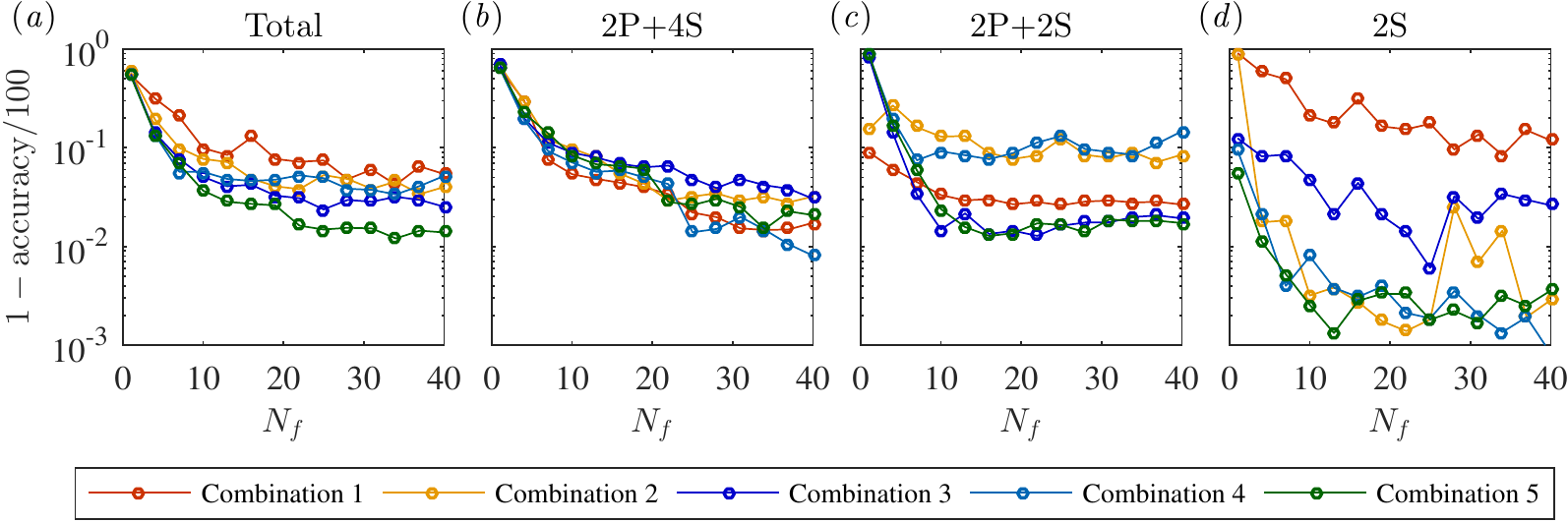}
\caption{%
Error incurred by five different trained networks in predicting the wake type or class of the test simulations. The error is plotted as a function of the number of features $N_f$ for (\textit{a})~all classes, as well as {for the individual} classes
(\textit{b})~2P+4S,
(\textit{c})~2P+2S, and
(\textit{d})~2S. 
}
\label{fig:errors_features}
\end{figure*}

Given the training dataset of inputs $\textbf{s}_m\in \textbf{S}$ and outputs 
$\textbf{y}_m \in \textbf{Y}$. The optimization problem can be written as
\begin{equation}
\bm{\uptheta}^\ast = \underset{\bm{\uptheta}}{\text{argmin}} \ \frac{1}{M} \sum_{m = 1}^M l_m( \textbf{y}_m, f(\textbf{s}_m,\bm{\uptheta})),
\label{eq:opt_problem}
\end{equation}
where $l_m$ is the loss associated with $\textbf{s}_m$ as in~\eqref{eq:loss} and
the total loss is an average of the losses over all snapshots in the training set.
To solve~\eqref{eq:opt_problem}, we use a stochastic gradient descent method,
\begin{equation}
\bm{\uptheta}^{\textrm{new}} = \bm{\uptheta}^{\textrm{old}} - \eta \frac{\partial}{\partial\bm{\uptheta}} \left( \frac{1}{M} \sum_{m} l(\textbf{y}_m,f(\textbf{s}_m,\bm{\uptheta}^{\textrm{old}}) \right),
\label{eq:grad_descent}
\end{equation}
where $\partial(\cdot)/\partial \bm{\uptheta}$ is the gradient of the loss function
and $\eta$ is a scalar called the \emph{learning rate}.
Equation~\eqref{eq:grad_descent} is applied iteratively until the change in loss reaches an acceptable threshold. 
In practice, in order to ensure that the solution to~\eqref{eq:opt_problem} is statistically significant, we need to have $N_p \ll M$, so that the problem is sufficiently overdetermined.

\definecolor{drkred}{rgb}{0.8,0.2,0.0}
\definecolor{orange}{rgb}{0.9,0.6,0.0}
\definecolor{drkblu}{rgb}{0.0,0.0,0.8}
\definecolor{midblu}{rgb}{0.0,0.4,0.7}
\definecolor{drkgrn}{rgb}{0.0,0.4,0.0}

\begin{table}[!t]
\centering
\begin{tabular}{c | c c c }
\textbf{Wake Type} &2P+4S &2P+2S &2S\\
\hline
{\color{drkred}\textbf{Combination 1}} &0.038 &0.065 & 0.079 \\
{\color{orange}\textbf{Combination 2}} &0.042 &0.067 & 0.073 \\
{\color{drkblu}\textbf{Combination 3}} &0.046 & 0.066 & 0.071 \\
{\color{midblu}\textbf{Combination4}} &0.040 & 0.063 & 0.077 \\
{\color{drkgrn}\textbf{Combination 5}}& 0.044 & 0.064 & 0.075\\

\end{tabular}
\caption{%
{The neural network is trained on five distinct combinations of training and testing data sets. In each combination,  four simulations from each wake type are selected for training and one simulation is reserved for testing.
The simulations reserved for testing are listed here.}
}
\label{tbl:sets}
\end{table}

\paragraph{Numerical implementation} We implement the neural network using Wolfram Mathematica version 11.1.1. 
The network is created using the {\texttt{NetChain}} function, where  the number of layers as well as $N$, $N_f$, $N_c$ are defined.
The linear layers are created using the {\texttt{LinearLayer}} function,  the activation layer using the {\texttt{ElementwiseLayer}} function with the {\texttt{Ramp}}  option, and  the normalization layer using {\texttt{SoftmaxLayer}}. 
This network is then trained using the {\texttt{NetTrain}} function and the training set $\textbf{S}$ together with its label set $\textbf{Y}$. 
After training, we evaluate the accuracy of the trained network on a distinct test dataset using the function {\texttt{ClassifierMeasurements}}.

\section{Results and Discussion}

We train the neural network on five distinct training sets obtained by arranging the fifteen flow simulations
into five combinations; for each wake type, four simulations are chosen for training and one simulation is reserved for testing. 
The simulations reserved for testing are highlighted in table~\ref{tbl:sets}.
In other words, in each combination, 80\% of the data was used for training and  the remaining 20\% was used to analyze the performance of the trained network. 

\begin{figure*}
\centering
\includegraphics{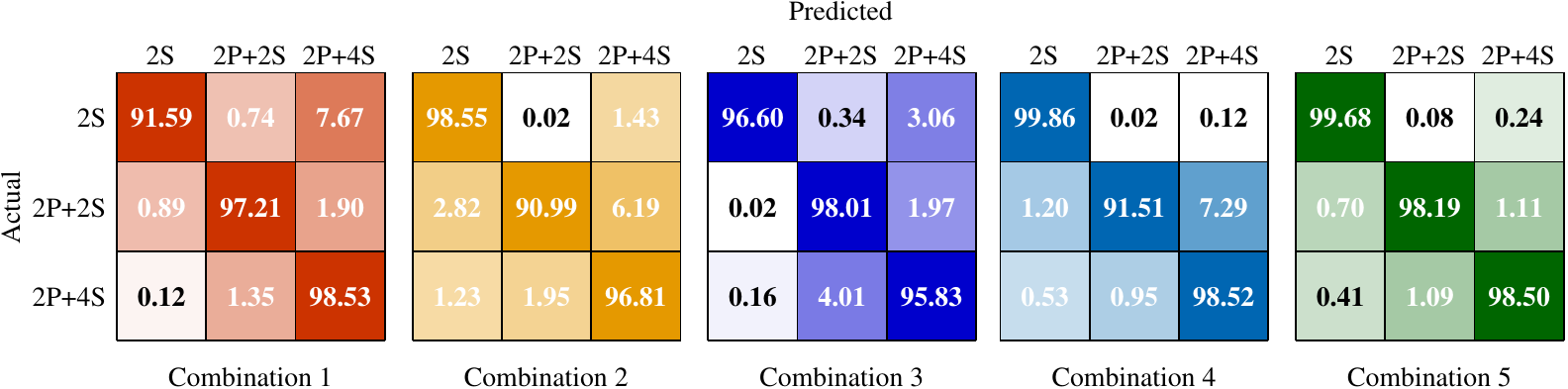}
\caption{Accuracy of the trained network in predicting the flow type of the test simulations. Diagonal entries show percentages of correct classification and off-diagonal entries show percentages of wrong classification and to which class they were erroneously assigned.}
\label{fig:confusion}
\end{figure*}

\paragraph{Network performance} We evaluate the performance of the trained network for each combination by measuring its accuracy in correctly classifying the data reserved for testing. 
Figure~\ref{fig:errors_features} shows the error $(=1-$accuracy$/100)$ in log-scale as a function of the number of features $N_f$ for all combinations.  The overall error for all classes is depicted in figure~\ref{fig:errors_features}(a) and the error per class in figures~\ref{fig:errors_features}(b)-(d). Combination 5 exhibits the best overall performance  for  $N_f \geq 10$, with errors as small as  0.01 and smallest error at $N_f = 34$. For comparison purposes, a purely random guess among three classes would incur an error of about $2/3$.

For the cases where the network fails, we track the fraction of cases assigned to each of the erroneous wake types or classes. The results are depicted in confusion matrices in figure~\ref{fig:confusion} for all combinations. The diagonal entries of these matrices show the accuracy of the network in correctly classifying to wake type and correspond to the results in figures~\ref{fig:errors_features}(b)-(d). The off-diagonal entries quantify the degree of confusion between classes in cases of erroneous classification. For example, the first row, third column of the confusion matrix shows the percentage of cases that belong to 2S but were misclassified as 2P+4S. Obviously, in all combinations and for all classes the accuracy of predicting the correct class is above 90\% and can be as high as 98 or 99\%. Note that higher degrees of confusion are observed between 2S, 2P+4S wakes and 2P+2S, 2P+4S wakes than between 2S, 2P+2S wakes. Also, in nearly every combination, there is a higher percentage of 2P+2S cases incorrectly classified as 2P+4S than 2S.
From a physical standpoint, the vortex filaments in 2S and 2P+2S wakes (see figure~\ref{fig:wakes}(c) and (d)) can be potentially confused with those of the 2P+4S wakes  (see figure~\ref{fig:wakes}(b)). Further, since 2P+2S and 2P+4S wakes are produced at decreasing frequencies of airfoil oscillations, the overall intensity of the vorticity field is lower, leading to potentially higher confusion between these wake types than with 2S wakes.

 \begin{figure*}
\centering
\includegraphics{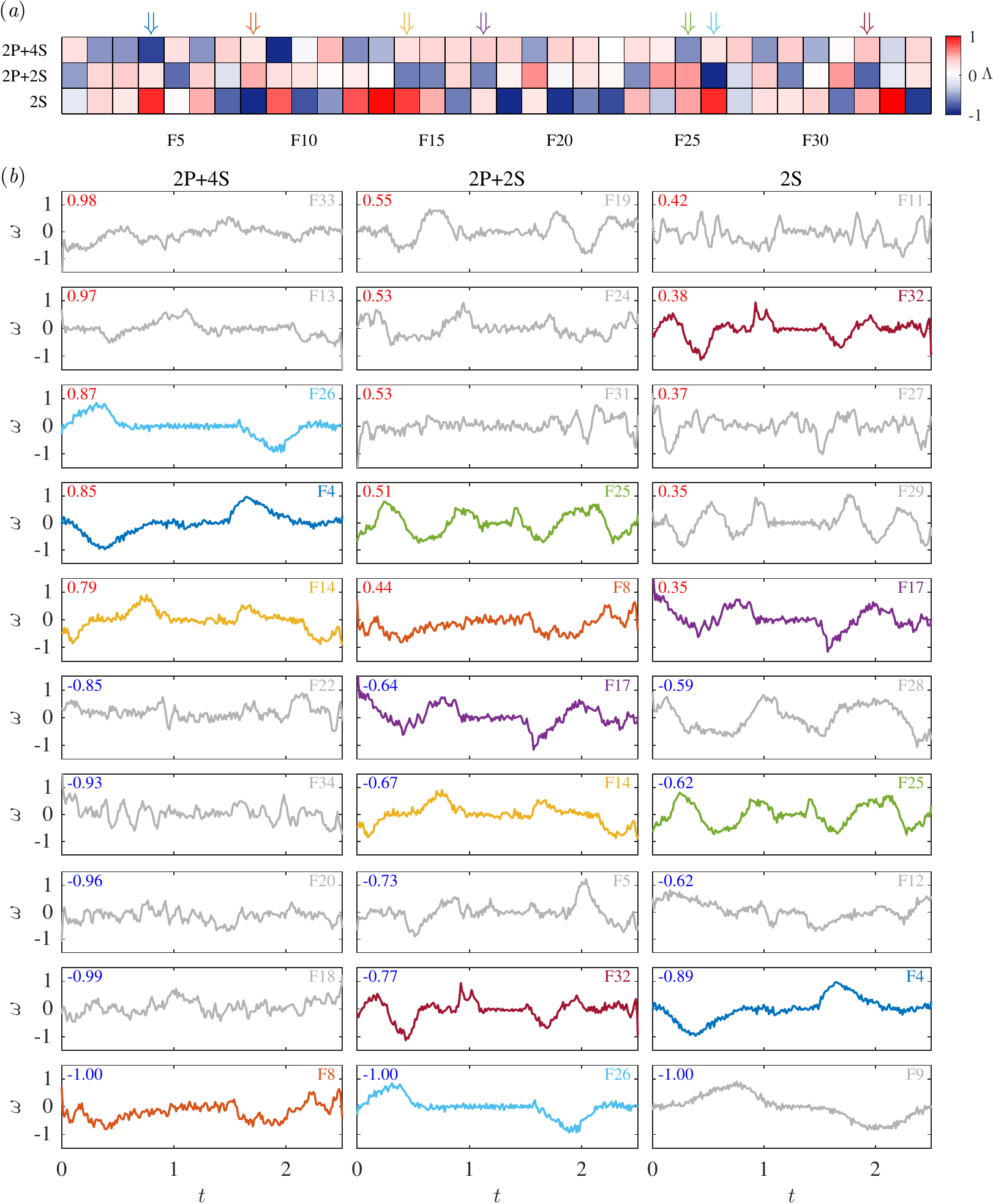}
\caption{%
(\textit{a})~Salience matrix $\Lambda$ of the $N_f = 34$ features of the model for $N_c = 3$ classes.
(\textit{b}) The five most positively and negatively salient features for each class are plotted as functions of time and ranked by salience.
 Features that occur as a pair of positive and negative identifiers for different classes are highlighted using the same color.
\label{fig:features}}
\end{figure*}

\paragraph{Salient features} To analyze the features relevant for differentiating between classes, we consider the trained network based on combination 5 and $N_f = 34$. 
The network features are the rows of $\textbf{W}$ as noted in~\S\ref{sec:network}. We refer to the $k^\textrm{th}$ row $\text{W}_{kn}$, $n=1,\ldots, N$, as feature $k$. 
Examples of the features arrived at by the trained network are plotted versus time in figure~\ref{fig:features}. 
While some features bear resemblance to the snapshots shown in figure~\ref{fig:time_traces}, most features 
are not recognizable by direct inspection of the vorticity time series. 

The first layer of the trained network projects a snapshot  onto the features of the network, thus quantifying the prevalence of each feature in the snapshot. 
A feature acts as either a positive or negative identifier. If a feature is a \emph{positive identifier} for a given class, its prevalence in a snapshot indicates a higher likelihood that the snapshot belongs to that class. Conversely,  if a feature is a \emph{negative identifier} for this class, its prevalence in a snapshot indicates 
a lower likelihood that the snapshot belongs to that class. 

The likelihood  $\hat{\text{y}}_c$ of  class $c$ is given by~\eqref{eq:layer_4}. Before normalization, its value is determined by 
\begin{equation}
\begin{split}
\text{Numerator}(\hat{\text{y}}_c) & = \exp\left(\sum \tilde{\text{W}}_{ck} \tilde{\text{q}}_k + \tilde{\text{b}}_c \right) \\
					 & = \prod_k \exp(\tilde{\text{W}}_{ck} \tilde{\text{q}}_k)\exp(\tilde{\text{b}}_c)
\label{eq:num}					 
\end{split}
\end{equation}
From~\eqref{eq:layer_1} and~\eqref{eq:layer_2}, we know that $\tilde{\text{q}}_k=\max(0,\text{W}_{kn}\text{s}_n + \text{b}_k)$ is zero or positive, reflecting which features are active in $\tilde{\text{q}}_k$. By substituting into~\eqref{eq:num},  it is straightforward to see that the weights $\tilde{\text{W}}_{ck}$ dictate how important  feature  $k$ is for  class $c$. With this understanding we define the \emph{salience} matrix $\Lambda_{ck}$,
\begin{equation}
\Lambda_{ck} = \frac{\tilde{\text{W}}_{ck} \sqrt{\sum_n \text{W}_{kn}^2}}{\underset{k}{\text{max}}\left( \left|\tilde{\text{W}}_{ck}  \sqrt{\sum_n \text{W}_{kn}^2}\right| \right)} ,
\end{equation}
where $\Lambda_{ck} \in [-1,1]$ and $\sqrt{\sum_n \text{W}_{kn}^2}$ is the $L^2$-norm of feature $k$. For similar measures of salience, see~\cite{Belue1995}.
The saliency matrix of combination 5 is shown in figure~\ref{fig:features}(a).

In figure~\ref{fig:features}(b), we rank the features by their salience and explore which features are most salient for each class.
In particular, we plot ten features per class: five corresponding to highest positive salience and five to lowest negative salience.
Some features such as F25 or F32 can be physically interpreted as counting the number of vortices passing over the vorticity sensor, but others do not lend 
themselves to such intuitive interpretation. 
Further, features that are most positively salient for one class are also most negatively salient for another class.
This duality reinforces the notion that features act as positive and negative identifiers; if a feature is very important for identifying 
a snapshot as belonging to a certain class, it is also likely that it is important for identifying the snapshot as \emph{not} belonging to the 
other classes. 
Figure~\ref{fig:features}(b) shows that for the class 2P+4S, most of the highly salient features, with salience values close to 1 and -1, do not appear with opposite level of salience in other classes, explaining why this class resulted in the most confusion. 

\begin{figure*}
\centering
\includegraphics{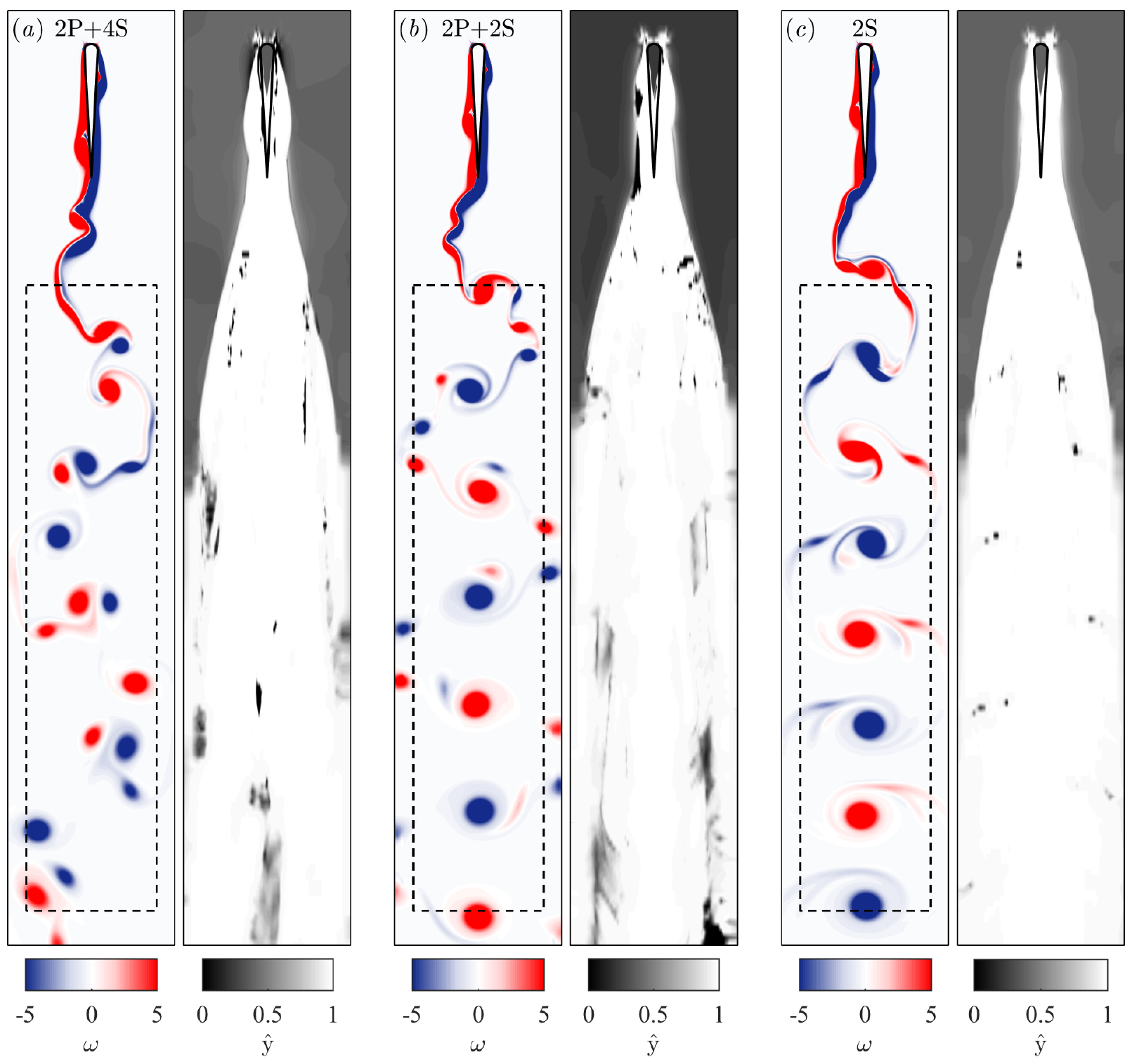}
\caption{%
Comparison of the vorticity field and the performance of the neural network in correctly classifying the wake as a function of the position $\bm{x}_m$ where the vorticity field is probed. (left) Contours of the vorticity field $\omega(\bm{x},t)$ repeated from figure~\ref{fig:wakes} to facilitate comparison and (right)  the spatial distribution of the likelihood $\hat{\text{y}}$ of correct classification for wake types
(\textit{a})~2P+4S at $St = 0.042$,
(\textit{b})~2P+2S at $St = 0.066$, and
(\textit{c})~2S at $St = 0.073$. 
}
\label{fig:wake_probabilities}
\end{figure*}

\paragraph{Mapping onto physical space} We examine the performance of the trained network on the wakes reserved for testing by mapping the network accuracy
onto the physical space. In figure~\ref{fig:wake_probabilities}, we plot the probability of correct classification as a function of space for the entire domain of the wake. In particular, we test on a larger region of the wake than the region we trained on, showing that the data that we used to train is somehow characteristic of the entire wake itself, thus supporting our hypothesis that local measurements contain enough information to classify the wake.
We also show that, save for some small areas downstream of the airfoil, the network has a very high probability of correct prediction, showing that the relevant information about vortical coherent structures is embedded in the time history of local vorticity throughout the wake.
The performance is degraded in the large regions (the dark conic-shaped zones around the airfoil) where there is essentially no information; the vorticity time series are always zero in these regions.

\section{Conclusions}

\emph{How} should one process local measurements to infer global information about the flow?
This question leads to a multi-valued inverse problem that is difficult to solve in general. 
Here,  we used neural networks to develop a processing algorithm that decodes the wake type from local vorticity measurements behind an oscillating airfoil.
One of the advantages of this data-driven approach is that it requires no a priori knowledge of the specific features of the decoding algorithm.
The patterns in the time series of the local vorticity (figure~\ref{fig:time_traces}) that are most useful for differentiating wake types are not readily recognizable by direct inspection.
Yet, the neural network uncovers the salient features of these wakes (figure~\ref{fig:features}) and maps a time series of the local vorticity to the overall wake type; the resulting mapping is both remarkably accurate and robust.

A few comments on the limitations of the feedforward neural network  employed here are in order. 
This simple network is not shift-independent. Namely, if a certain pattern occurs in a time series, the network will not recognize that it is the same pattern if it gets  shifted in time. To overcome this issue, convolutional neural networks will be used in future extensions of this work.
Another extension that would allow us to capture more complex relationships between input and output is to increase the number of hidden linear layers beyond one, creating a so-called deep neural network. 

We deliberately relied on simple local measurements, in the form of vorticity time series, in order to focus on the classification algorithm.
In the biological systems that motivated this work, the physiological modalities involved in flow sensing are more complex, with potentially distributed sensors tuned to the hydrodynamic cues relevant for the particular organism. 
These cues could be in the form of changes in the fluid velocity, pressure or density.
Flow sensors could independently or jointly probe one or more of these fields.
For example, the lateral line system of fish consists of two types of sensors that measure velocity~\cite{Kroese1992} and pressure~\cite{Coombs2005} along the fish body. Models of fluid-structure interactions in individual and schools of fish as well as behavior-based fish models often ignore this flow sensing ability; see, e.g.,~\cite{Becker2015, Fillela2017,Tsang2013} and references therein. A notable exception is our recent work that uses pressure sensors and postulates a behavior-based model coupled to a physics-based flow model to study fish rheotaxis, or alignment to an oncoming flow field,~\cite{Colvert2016}. 
It will be of major interest to train neural networks to detect flows using pressure and velocity sensors and to implement the trained networks in behavior- and physics-based models of fish swimming in isolation and in groups.

\paragraph{Acknowledgements}

We thank Haibo Dong and Geng Liu for providing access to their computational fluid solver and for their helpful assistance in setting up the flow simulations.
Funding for this work was provided by the Office of Naval Research (ONR) through grants N00014-14-1-0421 and N00014-17-1-2287 and the Army Research Office (ARO) through the grant W911NF-16-1-0074 (to E.K.). 
Brendan Colvert acknowledges support from the Department of Defense (DoD) through the National Defense Science \& Engineering Graduate Fellowship (NDSEG) Program.

\bibliographystyle{abbrv}
\bibliography{wake_learning}

\begin{thebibliography}{10}

\bibitem{Becker2015}
A.~D. Becker, H.~Masoud, J.~W. Newbolt, M.~Shelley, and L.~Ristroph.
\newblock Hydrodynamic schooling of flapping swimmers.
\newblock {\em Nature Communications}, 6, 2015.

\bibitem{Belue1995}
L.~M. Belue and K.~W. Bauer.
\newblock Determining input features for multilayer perceptrons.
\newblock {\em Neurocomputing}, 7(2):111--121, 1995.

\bibitem{Bozkurttas2009}
M.~Bozkurttas, R.~Mittal, H.~Dong, G.~Lauder, and P.~Madden.
\newblock Low-dimensional models and performance scaling of a highly deformable
  fish pectoral fin.
\newblock {\em Journal of Fluid Mechanics}, 631:311--342, 2009.

\bibitem{Bratt1953}
J.~B. Bratt.
\newblock Flow patterns in the wake of an oscillating aerofoil.
\newblock Technical report, Ministry of Supply -- Aeronautical Research
  Council, 1953.

\bibitem{Buchholz2008}
J.~H. Buchholz and A.~J. Smits.
\newblock The wake structure and thrust performance of a rigid low-aspect-ratio
  pitching panel.
\newblock {\em Journal of Fluid Mechanics}, 603:331--365, 2008.

\bibitem{Colvert2017b}
B.~Colvert, K.~K. Chen, and E.~Kanso.
\newblock Bioinspired sensory systems.
\newblock {\em Journal of Nonlinear Science}, 27:1183--1192, 2017.

\bibitem{Colvert2017a}
B.~Colvert, K.~K. Chen, and E.~Kanso.
\newblock Local flow characterization using bioinspired sensory information.
\newblock {\em Journal of Fluid Mechanics}, 818:366--381, 2017.

\bibitem{Colvert2016}
B.~Colvert and E.~Kanso.
\newblock Fishlike rheotaxis.
\newblock {\em Journal of Fluid Mechanics}, 793:656--666, 2016.

\bibitem{Coombs2005}
S.~Coombs and S.~Van~Netten.
\newblock The hydrodynamics and structural mechanics of the lateral line
  system.
\newblock In {\em Fish Biomechanics}, volume~23 of {\em Fish Physiology}, pages
  103--139. Academic Press, 2005.

\bibitem{Dahlkamp2006}
H.~Dahlkamp, A.~Kaehler, D.~Stavens, S.~Thrun, and G.~Bardski.
\newblock Self-supervised monocular road detection in desert terrain.
\newblock In {\em Robotics: Science and Systems}, volume~38. Philadelphia,
  2006.

\bibitem{Dehnhardt2001}
G.~Dehnhardt, B.~Mauck, W.~Hanke, and H.~Bleckmann.
\newblock Hydrodynamic trail-following in harbor seals (\emph{{P}hoca
  vitulina}).
\newblock {\em Science}, 293:102--104, 2001.

\bibitem{Engelmann2000}
J.~Engelmann, W.~Hanke, J.~Mogdans, and H.~Bleckmann.
\newblock Neurobiology: Hydrodynamic stimuli and the fish lateral line.
\newblock {\em Nature}, 408(6808):51--52, 11 2000.

\bibitem{Esteva2017}
A.~Esteva, B.~Kuprel, R.~A. Novoa, J.~Ko, S.~M. Swetter, H.~M. Blau, and
  S.~Thrun.
\newblock Dermatologist-level classification of skin cancer with deep neural
  networks.
\newblock {\em Nature}, 542(7639):115--118, 2017.

\bibitem{Fillela2017}
A.~Filella, F.~Nadal, C.~Sire, E.~Kanso, and C.~Eloy.
\newblock Hydrodynamic interactions influence fish collective behavior.
\newblock {\em arXiv:1705.07821}, 2107.

\bibitem{Gazzola2014}
M.~Gazzola, B.~Hejazialhosseini, and P.~Koumoutsakos.
\newblock Reinforcement learning and wavelet adapted vortex methods for
  simulations of self-propelled swimmers.
\newblock {\em SIAM Journal on Scientific Computing}, 36(3):B622--B639, 2014.

\bibitem{Gazzola2016}
M.~Gazzola, A.~A. Tchieu, D.~Alexeev, A.~de~Brauer, and P.~Koumoutsakos.
\newblock Learning to school in the presence of hydrodynamic interactions.
\newblock {\em Journal of Fluid Mechanics}, 789:726--749, 2016.

\bibitem{Godoy2008}
R.~Godoy-Diana, J.-L. Aider, and J.~E. Wesfreid.
\newblock Transitions in the wake of a flapping foil.
\newblock {\em Physical Review E}, 77(1):16308, 2008.

\bibitem{Haller2005}
G.~Haller.
\newblock An objective definition of a vortex.
\newblock {\em Journal of Fluid Mechanics}, 525:1--26, 2005.

\bibitem{Hecht-Nielsen1989}
R.~Hecht-Nielsen.
\newblock Theory of the backpropagation neural network.
\newblock In {\em International Joint Conference on Neural Networks}, pages
  593--605. IEEE, 1989.

\bibitem{Hornik1989}
K.~Hornik, M.~Stinchcombe, and H.~White.
\newblock Multilayer feedforward networks are universal approximators.
\newblock {\em Neural Networks}, 2:35--366, 1989.

\bibitem{Jeong1995}
J.~Jeong and F.~Hussain.
\newblock On the identification of a vortex.
\newblock {\em Journal of Fluid Mechanics}, 285:69--94, 1995.

\bibitem{Koochesfahani1986}
M.~Koochesfahani.
\newblock Wake of an oscillating airfoil.
\newblock {\em Physics of Fluids}, 29(9):2776--2776, 1986.

\bibitem{Koochesfahani1989}
M.~M. Koochesfahani.
\newblock Vortical patterns in the wake of an oscillating airfoil.
\newblock {\em AIAA Journal}, 27(9):1200--1205, 1989.

\bibitem{Kroese1992}
A.~B. Kroese and N.~A. Schellart.
\newblock Velocity- and acceleration-sensitive units in the trunk lateral line
  of the trout.
\newblock {\em Journal of Neurophysiology}, 68(6):2212--2221, 1992.

\bibitem{Lai1999}
J.~Lai and M.~Platzer.
\newblock Jet characteristics of a plunging airfoil.
\newblock {\em AIAA Journal}, 37(12):1529--1537, 1999.

\bibitem{Lee1997}
C.~Lee, J.~Kim, D.~Babcock, and R.~Goodman.
\newblock Application of neural networks to turbulence control for drag
  reduction.
\newblock {\em Physics of Fluids}, 9(6):1740--1747, 1997.

\bibitem{Mittal2008}
R.~Mittal, H.~Dong, M.~Bozkurttas, F.~Najjar, A.~Vargas, and A.~von Loebbecke.
\newblock A versatile sharp interface immersed boundary method for
  incompressible flows with complex boundaries.
\newblock {\em Journal of Computational Physics}, 227(10):4825--4852, 2008.

\bibitem{Montgomery1997}
J.~C. Montgomery, C.~F. Baker, and A.~G. Carton.
\newblock The lateral line can mediate rheotaxis in fish.
\newblock {\em Nature}, 389(6654):960--963, 1997.

\bibitem{Muller1999}
S.~M{\"u}ller, M.~Milano, and P.~Koumoutsakos.
\newblock Application of machine learning algorithms to flow modeling and
  optimization.
\newblock {\em Center for Turbulence Research Annual Research Briefs}, pages
  169--178, 1999.

\bibitem{Muller2008}
U.~K. M{\"u}ller, J.~G. van~den Boogaart, and J.~L. van Leeuwen.
\newblock Flow patterns of larval fish: undulatory swimming in the intermediate
  flow regime.
\newblock {\em Journal of Experimental Biology}, 211(2):196--205, 2008.

\bibitem{Novati2017}
G.~Novati, S.~Verma, D.~Alexeev, D.~Rossinelli, W.~M. van Rees, and
  P.~Koumoutsakos.
\newblock Synchronisation through learning for two self-propelled swimmers.
\newblock {\em Bioinspiration \& Biomimetics}, 12(3):036001, 2017.

\bibitem{Peskin1977}
C.~S. Peskin.
\newblock Numerical analysis of blood flow in the heart.
\newblock {\em Journal of Computational Physics}, 25(3):220--252, 1977.

\bibitem{Platzer2008}
M.~F. Platzer, K.~D. Jones, J.~Young, and J.~C. Lai.
\newblock Flapping-wing aerodynamics: progress and challenges.
\newblock {\em AIAA Journal}, 46(9):2136, 2008.

\bibitem{Ristroph2015}
L.~Ristroph, J.~C. Liao, and J.~Zhang.
\newblock Lateral line layout correlates with the differential hydrodynamic
  pressure on swimming fish.
\newblock {\em Physical Review Letters}, 114:018102, 2015.

\bibitem{Schmidhuber2015}
J.~Schmidhuber.
\newblock Deep learning in neural networks: An overview.
\newblock {\em Neural Networks}, 61:85--117, 2015.

\bibitem{Schnipper2009}
T.~Schnipper, A.~Andersen, and T.~Bohr.
\newblock Vortex wakes of a flapping foil.
\newblock {\em Journal of Fluid Mechanics}, 633:411--423, 2009.

\bibitem{Spedding2013}
G.~Spedding.
\newblock Wake signature detection.
\newblock {\em Annual Review of Fluid Mechanics}, 2013.

\bibitem{Stremler2014}
M.~A. Stremler and S.~Basu.
\newblock On point vortex models of exotic bluff body wakes.
\newblock {\em Fluid Dynamics Research}, 46(6):061410, 2014.

\bibitem{Tsang2013}
A.~C.~H. Tsang and E.~Kanso.
\newblock Dipole interactions in doubly periodic domains.
\newblock {\em Journal of Nonlinear Science}, 23(6):971--991, 2013.

\bibitem{Tytell2004}
E.~D. Tytell and G.~V. Lauder.
\newblock The hydrodynamics of eel swimming.
\newblock {\em Journal of Experimental Biology}, 207(11):1825--1841, 2004.

\bibitem{Vargas2008}
A.~Vargas, R.~Mittal, and H.~Dong.
\newblock A computational study of the aerodynamic performance of a dragonfly
  wing section in gliding flight.
\newblock {\em Bioinspiration \& Biomimetics}, 3(2):026004, 2008.

\bibitem{Werbos1990}
P.~Werbos.
\newblock Backpropagation through time: What it does and how to do it.
\newblock {\em Proceedings of the IEEE}, 78(19):1550--1560, 1990.

\bibitem{Williamson1988}
C.~H.~K. Williamson and A.~Roshko.
\newblock Vortex formation in the wake of an oscillating cylinder.
\newblock {\em Journal of Fluids and Structures}, 2:355--381, 1988.

\bibitem{Ysasi2011}
A.~Ysasi, E.~Kanso, and P.~K. Newton.
\newblock Wake structure of a deformable joukowski airfoil.
\newblock {\em Physica D: Nonlinear Phenomena}, 240(20):1574--1582, 2011.

\end{thebibliography}

\end{document}